\documentclass[conference]{IEEEtran}
\usepackage{todonotes}
\IEEEoverridecommandlockouts
\usepackage{cite}
\usepackage{amsmath,amssymb,amsfonts}
\usepackage{algorithmic}
\usepackage{graphicx}
\usepackage{textcomp}
\usepackage[utf8]{inputenc}
\usepackage{hyperref}
\usepackage[capitalise,nameinlink]{cleveref}

\usepackage{mathrsfs}
\usepackage{mathtools}
\usepackage{dsfont}
\usepackage{enumitem}
\usetikzlibrary{shapes}

\renewcommand{\epsilon}{\varepsilon}
\usepackage[caption=false, font=normalsize, labelfont=sf, textfont=sf]{subfig}

\date{\today}

\begin{document}

\title{A unified software framework for solving traffic assignment problems\\
\thanks{This work is supported in part by the Office of Science of the
 U.S.~Department of Energy under contract No.~DE-AC02-05CH11231.}
}
 
\author{\IEEEauthorblockN{Juliette Ugirumurera\IEEEauthorrefmark{1}, Gabriel Gomes\IEEEauthorrefmark{2},
Emily Porter\IEEEauthorrefmark{3}, Xiaoye S. Li\IEEEauthorrefmark{1}, and Alexandre Bayen\IEEEauthorrefmark{2}\IEEEauthorrefmark{4}\IEEEauthorrefmark{3}}\\
\IEEEauthorblockA{\IEEEauthorrefmark{1}Computational Research Division, Lawrence Berkeley Nationa Lab}
\IEEEauthorblockA{\IEEEauthorrefmark{2} Institute of Transportation Studies, University of California at Berkeley}
\IEEEauthorblockA{\IEEEauthorrefmark{3}Department of Civil and Environmental Engineering, University of California at Berkeley}
\IEEEauthorblockA{\IEEEauthorrefmark{4}Department of Electrical Engineering and Computer Science, University of California at Berkeley}}

\maketitle

\begin{abstract}
We describe a software framework for solving user equilibrium traffic assignment problems. The design is based on the formulation of the problem as a variational inequality. The software implements these as well as several numerical methods for find equilirbria. We compare the solutions obtained under several models: static, Merchant-Nemhauser, `CTM with instantaneous travel time', and `CTM with actual travel time'. Some important differences are demonstrated.

\end{abstract}


\section{Introduction}
\label{sec:intro}
The term \textit{traffic assignment} captures an array of problems concerning the distribution of traffic over a network. Traffic assignment problems arise in transportation planning applications, such as in the traditional ``four-step'' procedure, where it occupies the final step, following trip generation, trip distribution, and mode choice \cite{mcnally2007four}.  The fundamental principle that guides most traffic assignment models is Wardrop's first principle \cite{wardrop1952some}, which states that among available alternatives, drivers will select routes that minimize their travel times. Because the speed of traffic tends to decrease as the number of vehicles on the route increases, this principle leads to an iterated game in which drivers test new routes, day after day, until they converge to a Nash equilibrium, a.k.a. a \textit{user equilibrium}. 

Since the pioneering work of Wardrop, the mathematical and numerical study of traffic assignment has progressed slowly.
This is perhaps due to a historical scarcity of traffic data and network information, but also because of the complexity of traffic behavior. \cite{peeta} gives several examples in which problems could not be solved for networks of useful size. It is not surprising therefore that the problem has received renewed interest with recent increases in traffic and network data, and computational power. Also driving this renewal are concerns about the `unintended consequences' of the widespread use of routing apps \cite{traffic_apps}, as well as advances in autonomous driving technologies.

An investigation into the techniques of traffic assignment reveals a multitude of traffic flow models, numerical methods, and performance metrics. As is usually the case, the more detailed the model, the more expensive the computation. However it is not clear under which circumstances the additional effort (or investment in computational resources) is necessary. For example, although dynamic models that capture essential congestion phenomena are desirable, the computation (as well as network configuration effort) involved in solving dynamic traffic assignment (DTA) problems can be large. On the other hand, existing static planning methods cannot fully model traffic congestion because of their inability to represent queue formation and dissipation when traffic demand exceeds road capacities\cite{nie2010solving}. 

This paper presents a software framework for addressing such questions. The software is based on the formulation of traffic assignment as a variational inequality, described by Nagurney in \cite{nagurney2013network}. This is a very general formulation, as it requires only continuity of the traffic model. Hence, it covers a wide range of models: static and dynamic, macroscopic and microscopic, `analytical' and `simulation-based', etc. The software is modular and extensible, in the sense that it provides interfaces for incorporating new models, cost functions, and numerical solvers. 

There are many other programs, both commercial and shared, that solve traffic assignment equilibrium problems. Our software is most related to software tools for dynamic traffic assignment problems as described in Table \ref{tab:softwares}. More details about these software programs are also provided below.

\begin{table*}[htbp]
\centering
\begin{tabular} { | l | p{30mm} | p{30mm} | p{30mm} | p{30mm} | } 
	\hline
	\hline
	\textbf{Software} & \textbf{Flow Dynamics} & \textbf{Routing} & \textbf{Problem addresses} & \textbf{Algorithm} \\ \hline
	DTALite      & Mesoscopic agent based using queuing models & Time dependent shortest path & Dynamic user equilibrium & Approximate gradient method \\ \hline
    Dynameq      & Discrete event based  & Time dependent shortest path. Experienced travel time dynamic UE. No enroute route change.& Dynamic user equilibrium & 1) Method of successive averages, 2) heuristic adaptation of gradient based method  \\ \hline
    DynaMIT & Mesoscopic model based on deterministic queuing & Behavioral modeling based on utility theory: path size logit &  Converge to observed path flows & Extended Kalman Filter and variations \\ \hline 
    DynusT & Anisotropic mesoscopic model  & All or nothing assignment to shortest path iterates until time dependent dynamic UE is achieved & Dynamic user equilibrium & Adaptation of gradient projection in which step size is based on “relative gap”  \\ \hline 
    Integration & Microscopic agent based model using speed-space relationship  & Energy optimal routing & Dynamic user equilibrium & 1. Method of successive averages, 2. Feedback assignment, 3. Frank-Wolfe \\ \hline
    Aequilibrae & NA & All-or-nothing shortest path & Static user equilibrium & All-or-nothing \\ \hline
	\hline
\end{tabular}
\caption{Softwares for Dynamic User Equilibrium}
\label{tab:softwares}
\end{table*}

 \textit{DTALite} \cite{zhou2014dtalite} uses a mesoscopic agent based model with link flow dynamics based on Newell’s simplified kinematics wave model, a spatial queue model, and a point queue model. DTALite uses agent-based algorithm to calculate dynamic traffic equilibrium. \textit{Dynameq} uses a discrete event traffic flow model in contrast to the more typical discrete time based models. The model is based on car following, lane changing, and gap acceptance. Two solver algorithms are employed: the Method of Successive Averages and a heuristic adaptation of a gradient based method. \textit{DynaMIT} \cite{DynaMIT,ben2001dynamit} is a mesoscopic model based on speed-density relationships and queuing theory. Routing is based on a path size logit model. Joint estimation of unknown state space parameters (including OD flow, route choice model parameters, traffic dynamic model parameters, and segment capacities) is conducted using an Extended Kalman Filter. \textit{DynusT} \cite{chiu2011dynust} uses the “anisotropic mesoscopic model” for flow dynamics. This is similar to a stimulus-response type of car following model in which the two defining features are 1) a vehicle’s speed is influenced by vehicles in front of it (in same lane and adjacent lanes), and 2) the influence of downstream traffic decreases as distance increases. The solution algorithm is based on an adaptation of gradient projection in which the step size is based on “relative gap” which is a measure of the difference between experienced travel time and time dependent shortest path travel time. \textit{INTEGRATION} \cite{rakha2012integration} uses a microscopic agent based flow model based on speed-spacing relationships, a speed differential between a vehicle and the vehicle immediately ahead of it, and an acceleration model. The authors jointly consider routing and solution algorithms. Their software incorporates: 1) time dependent method of successive averages, 2) time dependent sub-population feedback assignment, 3) time dependent agent feedback assignment, 4) time dependent dynamic traffic assignment, 5) time dependent Frank-Wolfe algorithm, 6) time dependent external routing, and 7) distance based routing.  There are also some static traffic assignment softwares such as \textit{Aequilibrae} which does not incorporate flow dynamics but rather solves for static user equilibrium. 

A variety of other dynamic traffic simulators exist (including DynaSMART, Aimsun, Sumo, etc.) although these software tools are generally used for traffic simulation as opposed to explicitly solving dynamic user equilibrium problems. 

In contrast to these, the software described here does not prescribe the model, cost-function, or numerical method, but rather serves as a platform for comparing different options, both in terms of computation and the solutions they produce. The software presently includes a static model, two dynamic models: the cell-transmission model and the Merchant-Nemhauser model, a travel time based cost-function, and three numerical algorithms: the Frank-Wolfe algorithm, the Method of Successive Averages, and the Extra-Projection Method. The code can be found here \cite{ta_solver}. 

\section{Problem formulation}
\label{sec:formulation}
The goal of the problem is to find a demand assignment that produces an equilibrium state trajectory in the sense of Wardrop\cite{wardrop1952some}. The traffic network  is represented as a graph. Vehicles enter and exit the network through origin and destination nodes. They travel through the network following \textit{paths}. 

\vspace{1em}

\begin{tabular}{ll}
$w\in\mathcal{W}$ & all origin-destination (OD) pairs. \\ 
$d_w : [0,T]\rightarrow \mathbb{R}^+ $  & Demand for OD pair $w\in\mathcal{W}$. \\ 
$\mathcal{P}_w$ & Set of available paths for $w\in\mathcal{W}$. \\
$\mathcal{P}=\{ \mathcal{P}_w \}_{w\in \mathcal{W}}$ & All paths. \\ 
$h_p : [0,T]\rightarrow \mathbb{R}^+ $ & Demand on path $p$. \\
$h=\{h_p | p\in\mathcal{P}\}$ : & Demand assignment.
\end{tabular}

\vspace{1em}

The demand $d_w$ for an OD pair $w\in\mathcal{W}$ is, in general, a function of time over $[0,T]$. Each of the vehicles in the demand profile $d_w$ will upon entry to the network, choose (or be assigned) a path from the set of available paths for its OD pair. The number of vehicles assigned to path $p$ is denoted with $h_p$, and is also a function of time. A \textit{demand assignment} is a collection of profiles $h_p$ for each path $p\in\mathcal{P}$. A demand assignment is \textit{feasible} if all of its entries are positive, and it accounts for all of the OD demand. 
\begin{align}
h_p(t) &\geq 0 & \forall p\in\mathcal{P},\;\forall t\in[0,T] \\
\sum_{p\in \mathcal{P}_w} h_p(t) &= d_w(t) & \forall w\in\mathcal{W}  ,\;\forall t\in[0,T] 
\end{align}
Denote the set of feasible demand assignments with $\mathcal{H}$. 

A demand assignment produces, through the traffic dynamics, a cost profile $c_p(t)$ for each path $p\in\mathcal{P}$. 
A feasible demand assignment $h$ is also an \textit{equilibrium} assignment if its induced cost satisfies Wardrop's first principle: 

\noindent For each $w\in\mathcal{W}$ and $p\in \mathcal{P}_w$,
\begin{equation}
h_p(t) > 0 \quad
 \Rightarrow \quad c_p(t) \leq c_{p'}(t) \quad \forall p'\in\mathcal{P}_w \;,\; 
 \forall t\in[0,T]
\end{equation}
We denote the map from demand assignments to path costs with $F : \mathcal{H}\rightarrow \mathcal{C}$. 
\begin{equation}
c = F(h)
\end{equation}
$c$ is the collection of path costs $c_p(t)$.

As proven in \cite{patriksson2015traffic}, a feasible assignment $h^*$ is an equilibrium assignment if and only if it satisfies the variational inequality,
\begin{equation}
\label{eq:vi}
\langle F(h^*), h-h^* \rangle \geq 0 \quad \forall h\in\mathcal{H}
\end{equation}
\cite{nagurney2013network} provides proofs of existence and uniqueness of solutions to (\ref{eq:vi}), respectively under conditions of continuity and strict monotonicity of $F$. It should be noted however that strict monotonicity is not generally to be expected of traffic, since the addition of a single vehicle to a nearly empty network may not affect the travel time of any single driver.

In the case that there exists a convex function $f$ such that $F=\nabla f$ ($F$ is monotone and $\nabla F$ is positive semi-definite), the problem can be posed as an optimization problem,
\begin{equation}
\label{eq:opt}
\begin{aligned}
& \underset{h}{\text{minimize}}
& & f(h) \\
& \text{subject to}
& & h \in \mathcal{H}
\end{aligned}
\end{equation}
Thus we can classify traffic assignment problems into two categories: those with positive semidefinite Jacobians, which are optimization problems, and the more general problems, which remain as variational inequalities. It is worth considering the phenomena that are lost in using path cost functions $F$ with positive semi-definite Jacobians. The symmetry of $\nabla F$ means that an additional unit of demand on any path $p$ has the same effect on another path $p'$ as an additional unit of demand on path $p'$ would have on $p$. This assumption precludes at least two important features of traffic models:
\begin{enumerate}
\item \textit{permissive left turns}: At intersections, vehicles are often allowed to turn left through gaps in oncoming traffic. When the flow of oncoming traffic is large, then these gaps are rare, and vehicles must wait longer to turn. However, because the oncoming traffic has the right-of-way, their travel times are not impacted by the number of vehicles waiting to turn left. 
\item \textit{backward propagation of congestion}: One of the essential features of traffic dynamics is that regions of high density propagate upstream. Consider two paths $p$ and $p'$ that cross at an intersection. Paths $p$ is nearly empty, but $p'$ is congested up to the intersection. An additional vehicle on $p'$ will block the intersection, and thus obstruct vehicles on $p$, but an additional vehicle on $p$ has no effect on $p'$.
\end{enumerate}
A second level of classification relates to whether the function $F$ is static or dynamic. Within optimization-based traffic assignment, these two types have naturally been treated using the techniques of mathematical programming and optimal control respectively. If the function $F(h)$ is monotone (strictly monotone), then $f(h)$ is convex (strictly convex).

\section{Traffic models}
\label{sec:models}
In this paper we will make use of three traffic models: a static model (ST), the Merchant-Nemhauser model (MN) and the cell-transmission model (CTM). The static model will employ a link-based cost function, and therefore will be solvable with standard convex optimization techniques. The MN model, although dynamic, does not propagate congestion. Its Jacobian matrix is symmetric and positive semidefinite, and hence it can be posed as an optimal control problem. The CTM replicates the backward propagation of congestion, and must be treated with the more general techniques of variational inequalities. 

\subsection{Static model (ST)}
Use $\mathcal{L}$ to denote the set of all links in the traffic network. Then, under the static traffic model, the flow on a link $\ell\in\mathcal{L}$ is the sum of all demands on paths that include link $\ell$. This information is gathered into an incidence matrix 
$\Delta\in\{0,1\}^{|\mathcal{L}|\times|\mathcal{P}|}$, whose $p$'th column has 1's in the positions of links in path $p$. For each time $t$, the network flow $f(t)\in\mathbb{R}^{|\mathcal{L}|}$ corresponding to a demand assignment $h(t)$ is computed with,
\begin{equation}
\label{eq:staticmodel}
f(t) = \Delta \; h(t) 
\end{equation}
The travel time on link $\ell$ is then computed as a function of the flow on link $\ell$, and the travel time on path $p$ as the sum of the travel times on the links that constitute path $p$,
\begin{equation}
\label{eq:statictt}
c_p(t) = \sum_{\ell\in p} \tau_\ell(f_\ell(t))
\end{equation}
The function $\tau_\ell(\cdot)$ is the flow-to-travel time function. The most widely used version of $\tau_\ell(\cdot)$ is the BPR function \cite{wiki_Route_Choice},
\begin{equation}
\label{eq:bpr}
\tau_\ell(f) = \tau^o_\ell\left( 1 + \gamma_\ell  \left( \frac{f}{\bar{f}_\ell} \right)^4 \right)
\end{equation}
$\tau^o_\ell$ and $\bar{f}_\ell$ are respectively the free-flow travel time and the capacity of link $\ell$. $\gamma_\ell$ is a tunable parameter, typically set to 0.15 \cite{wiki_Route_Choice}.
This is a convex optimization problem.


\subsection{Cell-transmission model (CTM)}
The CTM was introduced by Daganzo in \cite{daganzo1995cell}. 
Here we describe the application of the CTM approach to the path-based setup of our software. The state of the model is the number of vehicles in each link, segregated by path: $x_{\ell,p}(t)$. The state evolves according to,
\begin{equation}
x_{\ell,p}(t+\Delta t) \;=\; x_{\ell,p}(t) \;+\; f^{in}_{\ell,p}(t) \;-\; f^{out}_{\ell,p}(t)
\end{equation}
where $\Delta t$ is the time step. The computation of the incoming and outgoing flows for the links, $f^{in}_{\ell,p}(t)$ and $f^{out}_{\ell,p}(t)$, involve intermediate quantities: the link supplies $s_\ell(t)$ and link/path demands $d_{\ell,p}(t)$.
\begin{align}
s_\ell(t) &= S_\ell\left(\sum_{p} x_{\ell,p}(t)\right) \\
d_{\ell,p}(t) &= D_\ell(x_{\ell,p}(t))
\end{align}
$S_\ell(\cdot)$ and $D_\ell(\cdot)$ are respectively decreasing and increasing functions of their arguments. The incoming and outgoing flows are computed according to a \textit{node function} which takes as arguments, for each node, the demands and supplies of all links incident on the node. There are several alternatives for the node function, e.g.  \cite{wright2017node,tampere2011generic}. These differ in the general case, but in the one-to-one case reduce to the original CTM formulation in which the total flow through the node is given by:
\begin{equation}
\label{eq:f}
f(t) = \min\left( \sum_{p}d_{\ell',p}(t) \; , \; s_{\ell''}(t)  \right)
\end{equation}
$\ell'$ is the upstream link and $\ell''$ is the downstream link. This total flow is then apportioned to the different paths according to an assumption of uniform speed (or FIFO),
\begin{equation}
f^{out}_{\ell',p}(t) = f^{in}_{\ell'',p}(t) = f(t)\frac{x_{\ell',p}(t)}{x_{\ell'}(t)}
\end{equation}
$x_{\ell}(t) = \sum_{p} x_{\ell,p}(t)$ is the total number of vehicles in link $\ell$.

\subsection{Merchant-Nemhauser model (MN)}
In a seminal paper for DTA \cite{merchant1978model, merchant1978optimality}, Merchant and Nemhauser introduced a dynamical model for analyzing traffic assignment problems. The MN model can be understood as a special case of the CTM, in which the supply function $d_\ell(\cdot)$ is set to a constant value equal to the link capacity. The calculation of node flow of Eq. (\ref{eq:f}) then depends only on the demands in upstream links, $d_{\ell',p}(t)$, hence backward propagation of information is not possible, and vehicles accumulate without bound in bottleneck links.

\subsection{Travel time calculation for the CTM and MN models}
The path cost function $c_p(t)$ represents the travel time for a vehicle departing at time $t$ and traveling along path $p$ to its destination. It is calculated by accumulating the travel time on each link in path $p$. Two types of travel time calculation can be used: \textit{instantaneous} travel time and \textit{actual} travel time. The `instantaneous travel time on path $p$', $c_p^{inst}(t)$ is:
\begin{equation}
c_p^{inst}(t) = \sum_{\ell\in p} \tau_\ell(t)
\end{equation}
Here $\tau_\ell(t)$ is the travel time on link $\ell$ at time $t$. The `actual travel time' $c_p^{act}(t)$ is obtained by sequential accumulation of the travel times for each of the links in the path, each shifted by its travel time.
\begin{equation}
t^{enter}_{n(\ell)} = t^{exit}_{\ell} = t^{enter}_{\ell} + \tau_{\ell}(t^{enter}_{\ell})
\end{equation}
$n(\ell)$ is the link following $\ell$ along path $p$.
$c_p^{act}(t)$ is the exit time for the last link in the path.
We compute $\tau_\ell(t)$ with $\tau_\ell(t) = \Delta t \; \frac{x_\ell(t)}{f_\ell(t)}$. The technique of \cite{newell1993simplified} is also available, but was not used in this paper.

Two approaches are common for computing $\tau_\ell(t)$. The first relies on a definition of speed in the CTM as flow / density, and travel time as distance / speed. Considering units, one obtains,
\begin{equation}
\tau_\ell(t) = \Delta t \; \frac{x_\ell(t)}{f_\ell(t)}
\end{equation}
This formula has the caveat that $\tau_\ell(t)$ equals the free-flow travel time if either $x_\ell(t)$ or $f_\ell(t)$ equal zero. The second method, introduced by \cite{newell1993simplified}, uses the cumulative flow at the boundaries of the link.

\section{Algorithms}
\label{sec:algorithms}
\begin{figure}[h]
    \centering
    \includegraphics[width=0.4\linewidth]{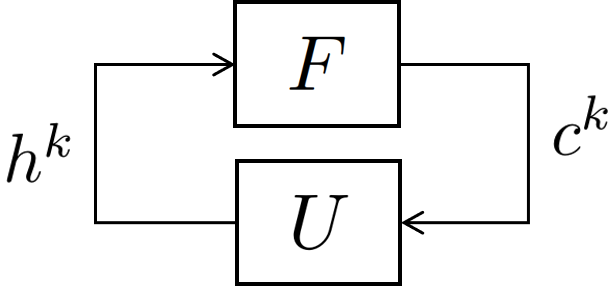}
    \caption{Generic iteration}
    \label{fig:iteartion}
\end{figure}

The experiments of Section~\ref{sec:experiments} involve several numerical algorithms for solving traffic assignment problems. These all follow the fixed-point iteration depicted in Figure \ref{fig:iteartion} between the model $F$ and the update function of the numerical method, $U$. $U$ is allowed to have a state. 
The numerical algorithms use the
\textit{all-or-nothing assignment at iteration $k$},  $y^k$:
\begin{equation}
\label{eq:allornothing}
y^k_p = \left\{
\begin{tabular}{ll}
$d_w/s$ & $p\in\underset{p'\in\mathcal{P}_w}{\text{argmin }} c^k_{p'} $ \\
0 & otherwise
\end{tabular}
\right.
\end{equation}
Here $s=|\underset{p\in\mathcal{P}_w}{\text{argmin }} c^k_{p}|$. The superscript $k$ is the iteration index of the numerical solver.
The all-or-nothing assignment is also used in the  termination criterion ($\epsilon>0$):
\begin{equation}
\label{termination_criterion}
\text{Stop if }
{\frac {\langle c^k,y^k-h^k \rangle} {\langle y^k, c^k\rangle}} \leq \epsilon
\end{equation}

\subsection{Frank-Wolfe Algorithm (FW)}
FW 
\cite{fukushima1984modified} 
is a well-known method for solving convex optimization problems which is especially well suited for network problems. The algorithm was used in \cite{thai2016negative} to solve large-scale static traffic assignment problems. The update function of FW is,
\begin{equation}
h^{k+1} = h^k +\alpha\; (y^k - h^k)
\end{equation}
with $\alpha$ set such that $h^{k+1}$ minimized the cost function over the chord between 
$y^k$ and $h^k$.



\subsection{Method of Successive Averages}
The Method of Successive Averages (MSA) is a heuristic that does not guarantee convergence to a solution, but has generally been found to work well \cite{nie2010solving}. The algorithm advances with,
\begin{equation}
h^{k+1} = (1-1/k)h^k + (1/k) y^k
\end{equation}


\subsection{Extra Projection Method}
The Extra Projection Method (EPM) is based on the Euclidean projection operator, defined as,
\begin{equation}
\Pi_\mathcal{H}(x) = \underset{h}{\text{argmin}}\{\lVert h-x\rVert_2 \; : \;h \in\mathcal{H} \}
\end{equation}
The EPM guarantees convergence when $F$ is Lipschitz continuous and pseudo-monotone \cite{nie2010solving}. 
$\tau^k$ is a number that is smaller than the Lipschitz constant of $F$. The update function of EPM is,
\begin{equation}
h^{k+1} = \Pi_\mathcal{H}(h^k - \tau^k F(z^k))
\end{equation}
where $z^k = \Pi_\mathcal{H}(h^k - \tau^k c^k)$. If the Lipschitz constant of $F$ is unknown, then \cite{nie2010solving} proposes the following update equation for $\tau^k$:
\begin{equation}
\tau^{k+1} = \left\{
\begin{tabular}{ll}
$\sigma\;\tau^k$ & 
if $y^{k+1}-y^k < 0$ and $\frac{|y^{k+1}-y^k|}{|y^k|}> \mu$ \\
$\tau^k$ & otherwise
\end{tabular}
\right.
\end{equation}
$\mu$ and $\sigma$ are scalars between 0 and 1.

\section{Software}
\label{sec:software}
\begin{figure}[h]
    \centering
    \includegraphics[width=0.7\linewidth]{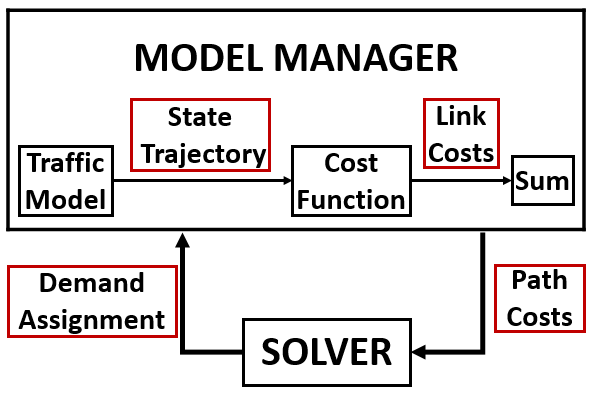}
    \caption{Software design.}
    \label{fig:Block_Diagram}
\end{figure}

Using the VI formulation and the generic iteration shown in Figure \ref{fig:iteartion}, we designed a modular and extensible software framework to solve traffic assignment problems, including static and dynamic traffic problems. Figure~\ref{fig:Block_Diagram} show that the software framework has two main modules: the Model Manager and the Solver modules. The Demand Assignment and Path Costs components are wrappers for demand assignment $h$, and for path travel costs respectively $c$.

The Model Manager has three components that correspond to function $F$: the Traffic Model, Cost Function and Sum components. The software framework currently include three traffic models: ST, MN, and CTM, which all integrate in the framework via the Traffic Model component. These are implemented using the BeATS simulator \cite{beats}, which is an open-source implementation of the dynamic traffic models resported in this paper.  The State Trajectory module is a wrapper for link states (e.g: flow per link), and the Link Cost module is a wrapper for the travel cost per link (e.g experienced travel time on a link). The Model Manager's role is to translate an assignment $h$, specified as a sequence of demand per path by the Demand Assignment module, into the corresponding path costs, represented as a sequence of cost per path.

The Solver module serves as an interface to plug in solution algorithms for DTA. The Solver works in a loop (as described in Figure \ref{fig:iteartion}) in which it generates candidate demand assignments, and expects to be given the corresponding network path costs. This loop continues until an equilibrium demand assignment is reached. The Solver interface allows to incorporate different DTA solution algorithm. We included three solver algorithms: the MSA, the FW, and the EPM, which can be applied depending on the $F$ function properties. 

The advantage of our modular software framework is that it can be easily extended to address different traffic assignment problems by including new traffic models, cost functions, and solver algorithms. For example, the framework has a travel time based cost function. A user can add an energy or emission based cost function. For problems, such as simulation-based DTA problems, where the traffic model is tightly coupled with the cost function evaluation, a user can integrate implementation of the $F$ function without having to write the Traffic model, Cost Function and Sum modules. The complete documentation on software framework, with installations instructions can be found at \cite{ta_solver}.

\section{Experiments}
\label{sec:experiments}
\begin{figure}[htbp]
    \centering
    \includegraphics[width=0.9\linewidth]{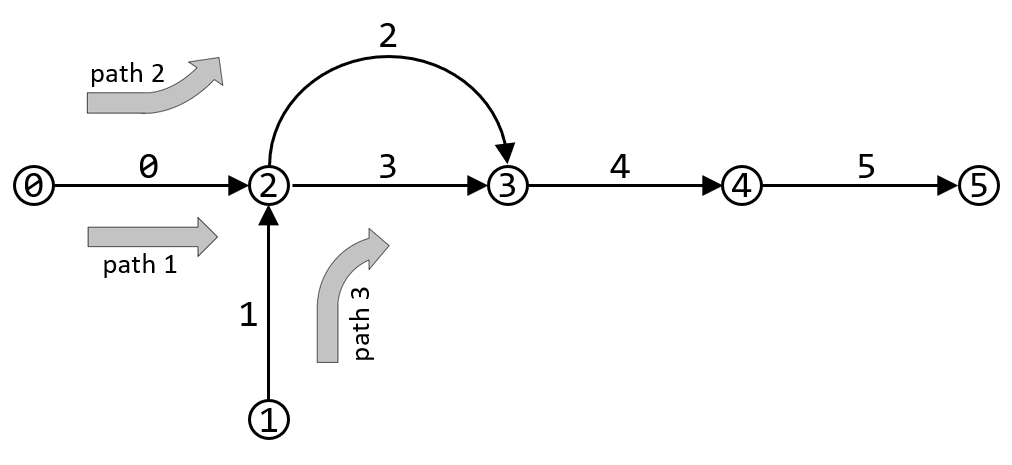}
    \caption{Traffic network.}
    \label{fig:config}
\end{figure}

In this section we demonstrate the methods described so far and implemented in the software using a simple scenario. The network, shown in Figure \ref{fig:config}, has six links numbered 0 through 5. Links 0 through 4 have flow capacities of 2,000 veh/hour, while link 5 has a capacity of 1,000 veh/hour. All links except for link 2 are 200 meters in length. Link 2 is 400 meters. The free-flow speed is 70 km/hr, and hence the free-flow travel time for all links except link 2 is about 10 seconds, while for link 2 it is about 20 seconds. There are two origin/destination pairs: OD 0, going from source node 0 to sink node 5, and OD 1, going from source node 1 to sink node 5. There are two paths available to OD 0: path 1, consisting of links [0, 3, 4, 5], and path 2 consisting of links [0, 2, 4, 5]. OD 1 is confined to path 3 = [1, 3, 4, 5]. Path [1, 2, 4, 5] is not utilized.

This network was chosen because it is small enough that the results are intuitive, and yet it illustrates the essential differences between the models and travel time functions described in Section~\ref{sec:models}. Only OD 0 has a choice of route, and hence the assignment problem is only to decide how the demand for OD 0 should be split between paths 1 and path 2 during each time interval. The total demand for both ODs will be set to a value that exceeds the capacity of link 5, and hence we expect congestion to form and propagate upstream past link 4, and onto to links 2 and/or 3, regardless of the assignment. Because path 1 is shorter than path 2, OD 0 should choose path 1 until the accumulated congestion on link 3 causes its travel time to be as large as that of link 2. Then the assignment should distribute traffic on paths 1 and 2 such that the travel times on links 2 and 3 are equalized. 

\subsection{Static assignment}
\begin{figure}[ht]
    \centering
    \includegraphics[width=0.8\linewidth]{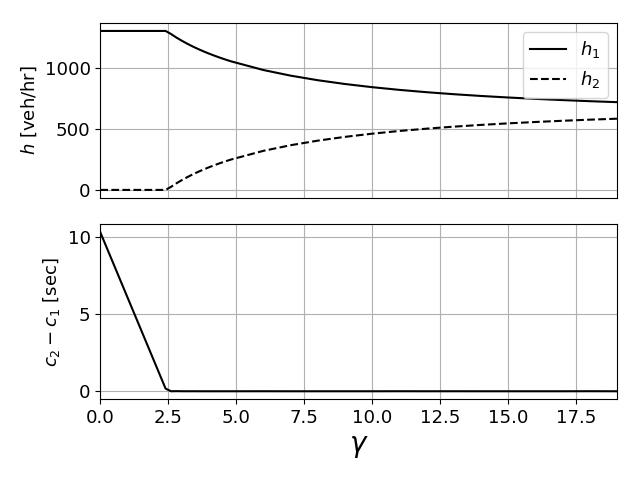}
    \caption{Static equilibria as a function of $\gamma$.}
    \label{fig:static}
\end{figure}

The static assignment problem with BPR cost function is the simplest case. It can be posed as a convex optimization problem and solved efficiently with the Frank-Wolfe procedure. Figure~\ref{fig:static} shows solutions obtained with OD demands set to $d_0$=1300 veh/hr, $d_1$=300 veh/hr, and $\gamma$ ranging from 0 to 20. Equilibria with $h_1\neq 0$ and $h_2\neq 0$ are characterized by $\tau_2=\tau_3$. Using Eq. (\ref{eq:bpr}), feasibility ($h_1+h_2=d_0$ and $h_3=d_1$), and $\tau^o_2=2\;\tau^o_3$, this leads to,
\begin{equation}
\bar{f}^4 + \gamma\left(\;2(d_0-h_1)^4 - (d_1+h_1)^4 \;\right) = 0
\end{equation}
Here $\gamma$ is the parameter of the BPR function, which has been assumed equal for all links. 
In the limit as $\gamma\rightarrow\infty$, the equilibrium value of $h_1$ decreases to 569.14 (the smallest real root of $2(d_0-h_1)^4 - (d_1+h_1)^4$), and $h_2=1300-h_1$ increases to 730.86. Thus, paradoxically, for large $\gamma$ the number of vehicles that choose path 2 exceed those that choose path 1, even though path 2 is longer than path 1. 

\subsection{Dynamic assignment}
\begin{figure}[ht]
    \centering
    \includegraphics[width=0.9\linewidth]{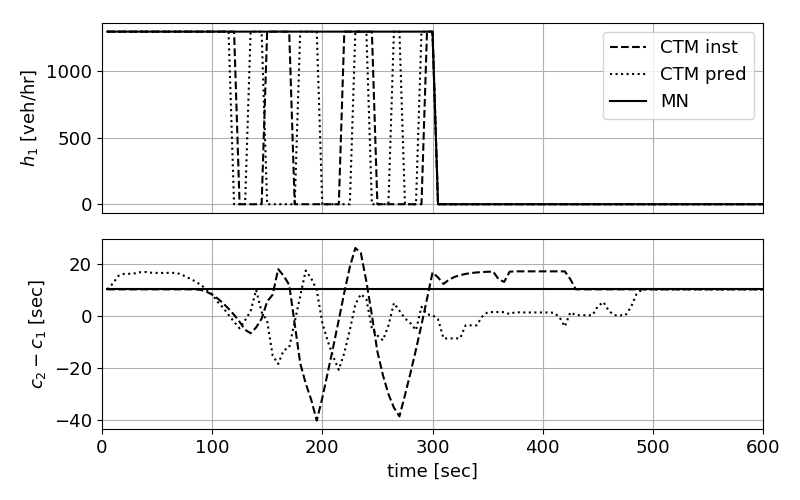}
    \caption{Dynamic equilibria - Assignment and cost trajectories.}
    \label{fig:ctm_vs_mn_hc}
\end{figure}

\begin{figure}[ht]
    \centering
    \includegraphics[width=0.9\linewidth]{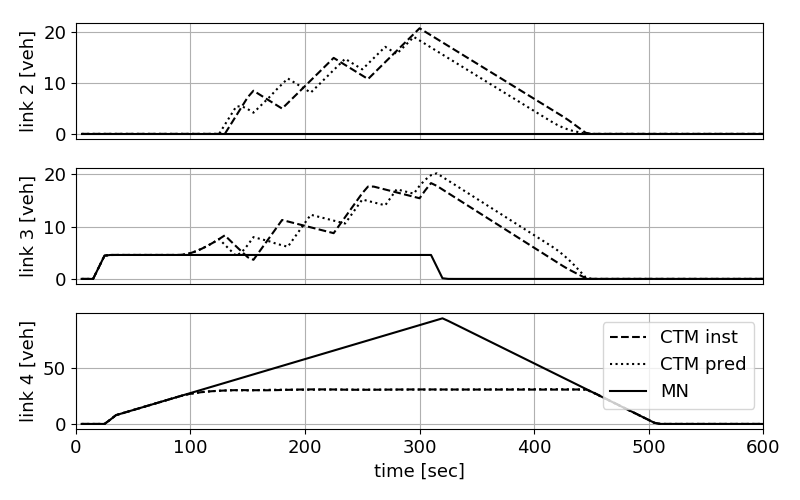}
    \caption{Dynamic equilibria - State trajectories.}
    \label{fig:ctm_vs_mn_x}
\end{figure}

Next we compute the dynamic equilibrium assignments under three different models: the MN model, the CTM model with instantaneous travel time, and the CTM with predictive travel time. The time period is 600 seconds. Demands are set to $d_0$=1300 and $d_1$=300 for time before 300 seconds, and $d_0$=$d_1$=0 thereafter. EPM was used to compute the solution, with an initial guess provided by MSA. The three solutions are shown in Figure~\ref{fig:ctm_vs_mn_hc}. The optimal assignment under the MN model places all of $d_0$ on path 1 for all times before 300 seconds, and so link 2 remains unused. This is because MN does not propagate congestion upstream, and so link 3 never slows down. This can be seen in Figure~\ref{fig:ctm_vs_mn_x}, where link 4 accumulates vehicles without bound up to 300 seconds. 

The two CTM solutions display behavior that is more realistic. In both cases, the entire demand is initially sent on path 1, since $c_2-c_1$ is positive, as can be seen in Figure~\ref{fig:ctm_vs_mn_hc}. The congestion on link 4 spills to link 3 after about 100 seconds, and at this point instantaneous $c_2-c_1$ begins to decrease. Notice that the predictive version of travel time reacts before congestion reaches link 3. The travel times on paths 1 and 2 are equalized at around 116 seconds for CTM predictive, and 121 seconds for CTM instantaneous. At this point, the demand is shifted entirely to path 2. The effect of this change is delayed by the travel time on link 0. In the meantime, $c_1$ continues to increase, and hence $c_2-c_1$ becomes negative. After the pulse reaches links 2 and 3, the travel time on path 1 decreases rapidly, and travel time on path 2 increases. $c_2-c_1$ again becomes positive, and another oscillation begins. 
We found across a range of values for $d_0$, that the size of the oscillations produced with predictive travel time is uniformly smaller than with instantaneous travel time. 


\begin{figure}[ht]
    \centering
    \includegraphics[width=0.9\linewidth]{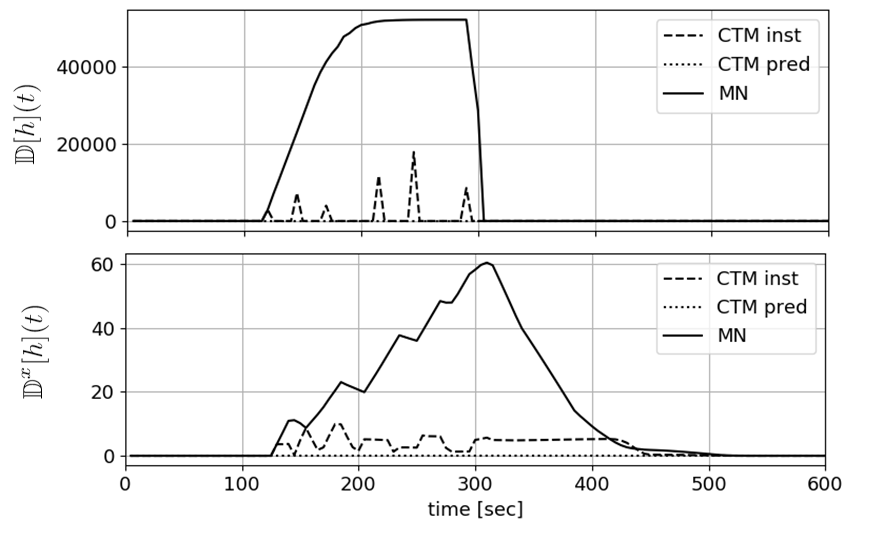}
    \caption{Two versions of `distance to Wardrop'.}
    \label{fig:dWardrop}
\end{figure}

Finally, we evaluate these three solutions in terms of their similarity to a Wardrop user equilibrium. To build a `distance to Wardrop' function, we express Wardrop's principle as a complementarity condition: $h^*$ is an equilibrium  if it is feasible and for all time $t$, all OD pairs $w$, and all paths $p\in\mathcal{P}_w$,
\begin{equation}
\label{eq:comp}
h_p(t)\left(c_p(t)-\pi_w(t)\right) = 0
\end{equation}
where $\pi_w(t)$ is the minimum travel time among paths in $\mathcal{P}_w$. The left hand side of Eq. (\ref{eq:comp}) is positive for all feasible assignments, and zero only for equilibrium assignments. Therefore the function,
\begin{equation}
\mathbb{D}[h](t) =\sum_w \sum_{p\in\mathcal{P}_w} h_p(t)\left(c_p(t)-\pi_w(t)\right)
\end{equation}
can serve as a rough measure of `distance to Wardrop', although it does not satisfy the triangle inequality nor the uniform scaling requirements of a norm. Computation of $\mathbb{D}[h](t)$ requires a model to provide the travel times $c_p(t)$ corresponding to $h$. Here we use the CTM with predictive travel time, since it is the model that most closely matches our intuitions about the outcomes of the experiment. The results are shown in Figure~\ref{fig:dWardrop}. Clearly the equilibrium due to the MN model is farther from a Wardrop equilibrium than either of the CTM results. A shortcoming of $\mathbb{D}[h](t)$ is that it only penalizes incorrect assignments $h(t)$ at time $t$. The lingering effects of vehicles sent on non-minimum paths are lost. If, however, the problem has a unique equilibrium solution, then the distance to that state trajectory can also be used to quantify `distance to Wardrop'.
\begin{equation}
\mathbb{D}^x[h](t) = \sum_{l} ||x^{*}(t)-x(t)||
\end{equation}
Here, $x^{*}(t)$ and $x(t)$ are, respectively, the equilibrium state trajectory and the state trajectory due to $h$. This quantity is depicted in the lower plot of Figure~\ref{fig:dWardrop}. Notice that the error persists beyond $t$=300.

\section{Conclusions}
\label{sec:conclusions}
We have presented a unifying software framework to solve traffic assignment problems based on a VI formulation. The software is not built to fit any particular traffic model; rather it has a modular design, which enables the solution and comparison of a range of traffic assignment problems. The two main modules, the model manager and the solver, provide interfaces to integrate different traffic models and numerical methods respectively. To demonstrate the software use, we conducted numerical experiments that compare equilibrium assignment resulting from three traffic models implemented in the framework: the static model, the MN model and the CTM model. The equilibrium assignments were calculated with one of the three numerical algorithms included: the Frank-Wolfe algorithm, the Method of Successive Averages, and Extra Projection Method. Results showed that, as expected, the equilibrium assignments from the MN and CTM dynamic models account for the traffic evolution better than the static model, which has no concept of queue formation on links. We also observed that the CTM model simulated congestion more accurately than MN, since it enables flow spill-back from links to upstream links as congestion grows.

In our future research, we will extend our framework for multiple commodities models to represent multiple driver classes such as app-routed and non-routed drivers. In addition, we observed that as the traffic network size grew to large-scale networks (a network with more than 2500 nodes and 2500 origin-destination nodes), the equilibrium computation time increased significantly (13 hours). Hence, we plan to continue exploiting distributed computation in HPC to speed up equilibrium calculations. This will enable the use of the framework for urban-scale traffic assignment problems and real-time traffic operations.

\bibliographystyle{IEEEtran}
\bibliography{citation.bib}

\end{document}